# 鉴于中方经验而在高等教育模式更新中匈牙利国家及大学领域的合作远景

玛卡依·奥迪拉·劳优世（Makai Attila Lajos）博士生，塞切尼·伊什特万大学（Széchenyi István University）（makai.attila.lajos@sze.hu）;拉姆哈普·邵博尔奇博士（Dr. Rámháp Szabolcs PhD）助理教授，塞切尼·伊什特万大学（Széchenyi István University）（ramhap@ga.sze.hu）

在匈牙利的高等教育模式更新不仅是大学与工业间的关系，及技术与知识传输过程的加深，而还在其领域创新政策成型中加强大学参与的重要性。此过程将为区域创新生态系统成员们之间的合作设立新框架，以及将经济-政府-学术系统的关系提升到更高层。在国家积极的参与方面、国家资源增持、及高等教育机构强力的创新组织者动力方面，无论是匈牙利还是中国的创新系统都是类同的。本论文试图总结中国的良好实践，如果能成功的学习，可有利于国家与大学之间的成功合作。通过研究的实践可将至今为止完成的大学模式更新过程放在一个新的环境中。本论文中主要通过杰尔市（Győr）的塞切尼·伊什特万大学而介绍其 过程。

《经济文献杂志》(JEL)编码：I23、I25、I28、O31
关键词：创业型大学 ；创新生态系统 ；三螺旋模型（Triple Helix）

# 绪言

匈牙利在2010年代有关高等教育最重要的政策变化就是公开支持变更为创业型大学模式的政府政策。此最后一步就是所谓的高等教育模式更新过程，其目标为建设有效的及现代化的高等教育机构，其过程中模式更新机构的创立及维护权力直接从国家手中转向信托基金会。这是走向创新能力开发及商业世界开放的重要一步。关注到较长时间的前情，在这个将近十年的过程中2020年八月1日起七所高等教育机构变更为新式运转是一个重要的里程碑。此变更的特殊点为：

— 机构从公共财政中提出，不再以财政预算机构方式而运营；

— 在公共财政以外，在最佳符合目标的基金会模式继续运转；

— 国家引入新的，客户逻辑融资系统（培训及绩效基础的融资），以及国家直接融资到基金会中；

— 结束机构中员工的公务员身份，全体改为普通员工身份。这些变化会带来更加灵活的运营：国家减少了20000-22000名的公务员数量；

— 国家的定额，按照协议方式，按财政法规每年交付给基金会。

按此逻辑为基础国家的任务为，向高等教育机构订购指定领域及数量的国家资助学生人数的订单，有关这方面的教学费用由国家承担，除此之外高校自己决定的，由国家接受后，自费方式完成教学（当然以每个学生为单位而计算的学校房地产运营费用也包括在其中）。除此之外，对于国家而言按照学府的学术性能力而在学术研究方面融资。高等教育机构的任务为按照国家指定的数量而培养专业人员以及完成学术性研究。同样重要的功能还有学府房地产的适当利用，非国家补助的，需交付学费的（越来越多的外籍）自费学生的培养，以及学术性，研发合作，市场/工业订单的履行，提供有关服务。对于机构而言是责任，但同时也是独立的，尽责的，以市场经济为基础开展资产管理实践的机遇(Rámháp, 2018)。

通过国家的指挥，但在区域基础上，由高等教育机构协调而完成的创新政策在世界上并不是独一无二的（匈牙利正走向这个方向）。这是一个在世界上多个国家都选择的模式，在此其中最脱颖而出的就是中国，早在1980年代就在中国中央指挥的同时已向以大学为中心的创新和区域性开发政策踏出了步伐。因此中国的实践和成果概况的了

解是重要的，通过认识，匈牙利决策方，机构可了解到此模式中的远景，以及可迎接创新政策和双边合作方向有可能的新机遇。

# 匈牙利和中国创新政策的发展弧线

在调查的两国中，就在地理位置和文化差异很大的情况下也可以在创新过程的历史中看到很多类同。对于两个国而言1949年都是一个重要的年份。1949年匈牙利象征性的接受苏联型宪法，结束了战争后的过渡期。中国的内战同样在1949年结束后成立了中华人民共和国。在两国形成的都是苏联式一党式政治系统。高等教育和文化开发领域同样形成的是苏联型模式（高等教育国有化，研发机构改革，由国家指挥的研发，匈牙利大学研发部门融入到匈牙利科学研究院）。在中国我们也看到了类似的过程，至1980年为止研究（研究院）和教育（大学）过程 是单独分开的，在邓小平主席提倡的改革后稍微有些变化，在高等教育层引入到研究和开发过程中及地方经济开发中（Chen等人，2016）。中国从1980年代起 无论是制度环境，还是研发所需的资源都不断的为在高等教育机构框架内形成研发结果的实用提供了越来越大的空间。主要根基为有关知识创造法规的接受，和1987年促进大学经营企业的成立（URI，高校校办产业），以及1993年接受的有 关高校知识财富以及知识产权的经营方面的法规（中国式拜杜法案）。

而匈牙利，则是到1989年为止，政权更迭为止研发领域主要由研究机构和 国有企业站为多数，高校领域主要工作为教育。教育与研发的隔离中仅在个别的研发-教育综合人士的职业道路上出现过例外。1989年后在国有企业私有 化后国有（学术）研发机构背后不在存在工业订购的支持，因此研发结果没有能够直接与工业和市场链轨。外籍跨国企业的出现也没有能够更改此现象， 因这些企业将生产，物流过程转移到匈牙利，但研究创新部门则保留在其总部或某些国际上重要的创新生态系统中。研发机构仅剩下了基础研究作用，但就是这样还是获得了多数重要的成就。对于匈牙利而言2004年加入欧盟给我们 带来了很大的改变，在此之后匈牙利机构（高校机构也在其中）也能够获得投标布鲁塞尔，以及NSRK（后来改为合作伙伴协议）框架内确保的国际等级的欧盟资源机遇。专业文摘评价中统一认同，从两千年初起我国的创新政策发展以及形成的外来力量主要来于匈牙利加入欧盟后；支持资源守规则的使用而出现的规范和结构不仅影响了资源分配方式，还影响到了整个有关政策机构体系(Dőry，2005)。尤其是欧盟结构资金欧盟规范体系背后的推动区域化和非集 权化的强大动力，也由匈牙利开发政策接受，将其政策列入即时的法规之中(Pálné Kovács，2013)。非集权化意向已经

在早期（接受区域化开发法规期间）就已出现，但其履行在入盟准备基金（PHARE,ISPA,SAPARD）执行时期，以及在第一届国家开发计划实践期间（2004-2006年）才开始实践。创新政策的计划和履行不仅由NSRK（后期改为合作伙伴协议）因欧盟基金规划和使用而成立，部分于政治无关的独立策略和组织性解决方法也有着很大的影响。在欧盟方不断的非集权化要求的同时，在匈牙利创新政策的过去20年中主要为非集权化和集权原则同时的出现和不断的替代。总而言之，一些欧盟财政期间不一定为统一的规定和组织框架。无论是2007-2013年，还是2014-2020年期间都发生过在运行周期期间发生了重要的及基础的原则变化，因此导致政策转向别方。创新政策在不同的策划期间有着不同的主要特征：

1、2007-2013年期间的特征为：

— 在创新政策中强调非集权化范例；

— 非集权化决议过程的分行和集权权益的干涉；

— 区域机构参与政策执行。

— 在非集权化机构体系中政策（及资助）决议过程的集权性质；

— 行政管理名义上使用区域性名称的计划和履行过程；其计划和履行非以创新生态系统上完成；

— 低强度技术转让过程，在创新政策中匈牙利科学院较多的，高校机构比较少的重要性。

— 缺少风险资金基础，招标资金为首要地位

2、2014-2020年期间的特征为：

— 集中化范例，国家经济层的重要性；

— 地方权益对集中化决议过程的干涉；

— 区域性创新机构被推及到后，或被淘汰；

— 政策（及资助）决策道路的集中化；

— 区域性创新生态系统名称开始被使用到各种不同的项目计划中和履行中；

— 技术转让过程，以及高等教育机构商业化观点的加强，匈牙利科学院继续为最主要的中心，以及高等教育机构在创新政策中加强区域性的角色；

— 国家，部分国家的（例如JEREMIE类）风险基金的出现。

中国在此期间。90年代中央政府越来越多次倡议以及越来越多的资金支持区域性研究高校的发展。在倡议中尤其重视名为《Project 985》及《Project 211》项目（Zhou,2019）。上述第一个项目从1998年启动，主要目标为加强中国高等教育系统在国际上的名气，以及在引入不同国家和区域资金后教育和研发基础设施的开发。第二个项目为1995年启动，为国家政府所倡议项目，目标为改善中国大学中的研发项目品质，以及将其研发嫁接到经济开发项目中。高校基础转让过程在两千年初带来了可视的结果，例如北京附近创新生态系统发展就出自其中。具有关调查（Zhou,2019年，第362页)结果可注意到高校基础转让业务带来了以下（对于区域经济也非常有关的）影响：

— 由高校给于有关新商业人员的培训及有关创业知识；

— 为已运转企业家提供专业级商业开发服务；

— 高校和工业方参与而实施的共同开发项目数量的增加；

— 研发业务所需基础设施及其它资源的分担；

— 通过大幅度授权而利用的IP市场；

— 崭新的，高校中心，技术密集型企业的成立。

周春彦在2007年研究调查中拿中国高等教育体系与西方高等教育体系相比其落后点（Zhou,2007）。调查指出以下几点问题：

— 高校机构人员：和发达国家相比培训程度落后，在发达国家评价高的高校学生所有教师都具备博士学位。

— 教育品质：两千年快速的发展导致教育没有能够同时升级。

— 研发能力：因乱序的本科教育而导致研发能力比较弱。

— 博士后奖学金人数数量稀少：为了增长其数字需要客观的资金来源，这样才有助于稳定的增长。

为解决上述问题中国高校教育系统无论在法规领域，还是资金方面都需要继续改革。在这方面需要提高中国高校国际化关系的主动性，而在作为全球高校排名基础的综合指标因素中（科学计量指标，高校知识产权的利用程度，与外国高校联合举办的本科

培训项目的数量）达到快速的和可观的成就。同样，融资结构的变更也被视为主要变化，其过程中在国家融资的同时出现私人资本，设立高校-私人资本及投资基金。北京大学内成立的方正集团（Founder Group）就是一个很好的例子，同样北京微生物科技有限公司（Peking University Biotech Co.），以及北京控股集团有限公司（Beijing Holdings Limited）也是好例子。国家-高校-私人领域的合作不仅解决了融资问题，还为那些期盼"大新闻"的股市投资商确保很好的投资机遇。最大的证明就是这些高校创办的以及由高校管理的高校校办产业（University Run Enterprise）上市以及股市的上升（Zhou，2019年，第373页）。类如北京大学校办的方正证券，获清华大学管理的同方股份或同方国信如今的价值都已经是刚上市时的好几倍。这也对匈牙利政策方面给予了证实，这是一个模范方式，为有创新开发倾向国通过加入更多资金投资而将资金翻倍的好办法。

## 创业型大学的理论背景

将传统的，以洪保德名义而创的双重高等教育范式（Etzkowitz，2019)边缘化后不仅使高校教育人数增长，而且高校技术转让业务的比例也会增长。通过在美国高校系统中将"科学作为资源（science as resource）"转换成"科学作为引擎(science as engine）"后强劲崛起的研发投资比例就可以看出使一个好的趋向(Popp Berman，2012)。服务型/创业型高校的成立同时也意味着经济逻辑融入到学术/高等教育系统中，而形成"学术资本主义"（Münch，2014)，其基础因素(尤其在高校排名中）无论在欧洲大陆，还是在英国或美国式高等教育体系中都促进经济逻辑的影响。在文摘中没有统一的创业型大学的名称，功能，独特点的解释。当然这还不证明在一些元素中没有获得共识；经济和学术体系之间的紧密关系，或以研发为中心从未被趋势或理论批评过。埃茨科威兹曾说："创业型大学是除了教育和研究大学使命外还促进经济/社会发展的具备第三使命的，越来越重要的当代现象，在此高等教育层作为基地总部承担生产中心组织及技术创新的角色"(Etzkowitz、Zhou,2018年，第58页）。有关这方面早期克拉克（1998年）（在部分大学模式隔离的同时）写下创业型大学的基础支柱为：专业管理，形成与大学有关的外设开发；融资多元化；通过商业观点而促进的学术背景；综合型创业文化。三螺旋模型框架中高校-经济-科学体系合作而形成高效益的创新首要条件，其中高校作为创新地点和整合者而确保综合因素（Leydesdorff、Etzkowitz，2001)，在此开放的创新平台上支持弥漫和互动创新过程的诞生（Chesbrough，2003)。此思维的进一步发展为卡拉雅尼斯（Carayannis）和坎贝尔（Campbell）2009年建立的四螺旋模型（

Quadruple Helix modell），这里第四"螺旋"是公众，以及当地的公众。同样两位作者在几年后认为作为创新最重要的推动力之一，作为其马达而是随着全球气候暖化所产生的环境及社会的挑战性。他们认为其重要性达到了以"第五螺旋"的身份而融入到创新模型中，因此诞生了五螺旋模型（Quintuple Helix modell）（Carayannis，2012)。

当然，高校作为在空间中呈现的元素，强劲的影响到较窄及较宽环境中的经济过程。查特顿（Chatterton）和戈达德（Goddard）在2000年通过模型而介绍高校及区域性经济的关系，主要强调相互依赖性质：一个强力的和创新的区域经济可支持高校的排名，以及一个有实力的高校可作为区域开发和创新的催化剂。查特顿-戈达德模型与三螺旋模型不同的是主要设立于区域化和分权化。模型将高校的角色和区域政治经济促进连接到了一起。在此概念中知识网络是区域经济发展主要元素，其中高校不仅是创新资源，还成为知识网络的交合点（Rámháp，2018年，第375页)。相关联的还要提到"学习型区域"(Hassink，2001)，为以知识为基础的区域发展的组织，文化和机构方面的分析框架。此范例在区域发展及创新领域中同样强调区域内高校机构的重要性(Rutten, 2003)。高校的区域角色，重要性，及强调分权化的模型和三螺旋模型都是彼此兼容的。一个模型强调的是高校机构在区域中的使命，另外一个则强调的是最佳运转结构，不同的分系统角色可为区域的创新生态系统的发展而最佳的合作。区域化，及区域开发政策因素和三螺旋模型可以集成并分析到其分析框架中（Ranga、Etzkowitz，2015）。我国的研究也获此肯定，确定在经济，文化和社会资源适当分配及区域性和政府角色按三螺旋（四螺旋）的合作而加强知识区域的设立机会（Rechnitzer，2016年，第245页）。

维塞玛（Wissema）（2009年）针对作为第三代高校功能标准的剑桥大学模型做出推论。全神贯注于一个自1970年代起一直不断的现象，因高校而促进剑桥附近形成了高新技术工业（剑桥现象）。维塞玛确定了七个主要元素：1、设定"第三个目标"：知识的利用及成为传播中心的角色；2、出席国际竞争市场成为主要目标；3、高校积极与工业及商业合作伙伴之间合作；4、研究性质的改变而出现跨学科和多学科研究；5、同时完成大众及精英培训；6、高校变成多文化机构；以及7、结束国家的直接融资和干涉。

东中欧在创新生态系统开发方面差一步速度的追随着欧洲中央国家的模型，当然在创新方面为区域边沿的状态在任何领域都使可见的。所谓的区域创新记分板（RIS）框架系统为基础计算的区域创新基数也证明了这一点。从2007年在所有领域都可以看到本区域和领先区域（例如：德国西部及南部，荷兰，英国南部，爱尔兰）相比还是比较

落后的（Gál、Páger，2018）。在高等教育中 还可注意到学生数量不断的减少（除了捷克和克罗地亚之外），以及在创新过程中国家的主要参与性。其它的特征就是区域创新生态系统中几乎不能够， 或仅有限的部分可以找到被视为三螺旋模型的有效运转所需要的RIO（区域创新的组织者），而使区域性创新，知识和共识的领域适当的组织，筹划和运行（Ranga、Etzkowitz，2015年，第124页)。在这个功能方面有越来越多比例由个别高等教育机构来代替区域政治不同的角色，在创业型大学变更后高校可完全适宜此任务。通过丽莎·妮特（Lisa Nieth）和保罗·本奈沃特（Paul Benneworth）（2019）的研究证明，一些边沿区域的发展需要部分高校在创新需求和供应之间的交流方面，确保支持基础设施方面，设立所需区域合作伙伴及所需能力开发方面的协助而完成。

# 中国高校环境中的经验：清华大学

1911年建校的清华大学毫无疑问的是中国最重要的教育机构。不仅因为其在 国际高校排名中所位于稳定的领先位置，而且在区域经济中的融入性及在全球创新生态系统中所承担的角色也是被肯定的。不可忽视的还有大学的历届毕业生中已经出了两名中华人民共和国主席（胡锦涛和习近平），这在在国家资本主义的基础上组织的中国体系中是一个很重要的资本。大学的使命宣言中宣布的目标中除了教育方面还有创意知识的开发，技术创新的协助，以及对社会目标贡献的目标在内，因此而包含了高等教育的三类使命。大学具备多种基础和应用研究部门，以及在全球也算是数一数二的技术转让功能，使大学和其机构成为国家及私人资金创新目标应用的主要角色。技术转让过程的组织和协调中主要任务由大学内以下机构完成（Zhou,2019年，第379页)：科技发展部门；产学合作委员会；海外项目部门；国际技术转移中心。

高级教育第三使命功能，及技术知识转让方面清华大学及中国高等教育体系的特点（与美欧体系不同的）为：

1、高校校办产业（URE）类大学衍生企业：

在欧洲高等教育体系中传统型衍生企业就是将所获得的知识财产在商业中传统式利用，而在美国（衍生活力很大的同时）主要在波斯顿附近及斯坦福大学附近形成的及在过去几十年全球越来越受欢迎的startup类逻辑的早期阶段企业来证 明大学的创新生态系统。传统衍生企业由大学导师/研究员成立，大学拥有少数股权（一般仅为由大学所有的知识产权许可协议）为其特征，大学和衍生企业在结构上，法律/规定上，财经和功能性上分割。Startup类运转模式中的创始方和 投资方之间所产生的投资协议中规范

其法律关系，在此大学的利益由企业中所有的大学股权（或有关知识产权的许可协议）一同在法律权力上和企业收入中，及转让退出中所获得分红。上述两个理想的运转模式与在中国而形成的结构完全不同，专业文摘称此模式为URE（高校校办产业）（Zhou,2019年，第369页)。其本质为将成立的衍生企业中具备知识产权的大学为多数股东，在此创业的导师/研究院在企业中为少数股东，其股权比例为传统的股东比例（5-10%），最多到30%为止。衍生企业和大学的管理没有完全的分割，大学的高校校办产业实际上融入到大学的管理系统之中。很多次企业的资金和预算也不会单独分割；实际上就是大学在没有中介机制的情况下直接步入经济分体系中的过程。因此在startup逻辑中自然而然的转让退出过程，作为投资商角度的目标在这个情况下根本不会产生：有大学所成立的URE仅因扩大及发展所需的资金投入的利益会接受私人投资资金，或上市（IPO），在肯定的大学控股或确保领导权力的情况下。成功的量角器在此不是最佳完成的转让退出，而是长期的稳定运转，其框架内有机会使由大学领导的URE自己也成为某工业内的投资商或收购商。在中国有多家重要的(如今已经成为具备全球供应链的跨国公司的）企业具备高校背景，成立时为高校领导，而且在企业获得跨国角色后还确保着大学的领导作用。

2、以高校为中心的投资基金：

高等教育范围活跃投资商的作用历史悠久。二十世纪中期在美国，传统的风险投资基金也是高校参与的，（波斯顿附近）大学创新生态系统框架内形成。不同规模的大学参与成为了所有全球主要投资生态系统的共同特征。在中国由国家领导的投资策略在当地，地区层，大学，地方政府杰尔地方（或国际）经济实体参与而形成。此体系的中央参与者为在地区层拥有主要影响的，由高等教育机构领导的/影响的投资基金，其积极部分无论是针对技术开发而需要的基础设施开发（例如：科技园），还是技术密集型早期阶段的企业投资。此投资基金在企业走向跨国市场时所需资源确保方面也具备同样重要的角色。

3、以高校为中心的创新生态系统国际化中大学与区域之间的合作：

高等教育机构之间国际合作无论在研究（共同的研发项目，合作而实践的，直接通过欧盟资金而实践的提议），还是在教育（双学位，联合学位培训教育，共同的大规模在线开放课程培训）功能方面的重要性越来越大。大学在第三使命方面看到的实践就比较少了。在此我们可见清华大学方面有一些值得重视的提议，UURR（大学-大学-区域-区域）模型（Zhou，2019年，第384页)。其重点为合作不仅在大学之间，而是高等

教育机构所在的区域之间，在合作中所有螺旋角色都参与。这样形成的合作不仅可协助教育及研究项目的协调，而且还打开了工业和经济实体合作的新型出口机遇，以及在部分项目融资方面为国家之间的合作形成新观点。实践样例为在中国的清华大学和浙江省，日本的岩手大学和岩手县协调合作的平台。而且大学和区域类似的项目在德国，以及千苏联多个区域都在进行。这种特殊的合作方式（创新整的，以及大学结构区别的不同）在美国或欧洲极少有。在MIT Sloan School协调的REAP（区域创业加速计划）项目框架中可见一种"半边"的合作。在此MIT加速器项目中有一区域大学，企业，大型企业，基金投资商，及政府共同参与，及共同学习由MIT开发的加速器方法，以及于MIT创新生态系统中的成员们设立合作关系。在此项目中从美方式没有区域政府方的参与，也没有共同投资，因此合作没有从教育合作中踏出脚步。

# 塞切尼·伊什特万大学成为创业型大学的过程

塞切尼·伊什特万大学式匈牙利多瑙河以西北部区域最大的高等教育机构。在此区域为国家最发达的农业和食品业领域，但工业才是这个区域的主要角色，从90年代后期开始不断的增长和发展。区域北部-尤其是杰尔-莫松-肖普朗州，和瓦世州一小部分区域-因地理位置活跃的介入欧洲及世界市场的血液循环中，比较优势的如今也是工业，尤其是机械和汽车业。此区域人均GDP超越了国家平均值（Tamándl，2014）。

大学的前身机构，交通与电信技术学院成立于1968年，其目标为在区域内 确保交通和电信基础设施开发及维护所需的技术知识分子。作为学院的机构在1986年改名为塞切尼·伊什特万。从1990年起除了工程师科学外还开始了健康科学，经济科学，和法律及艺术培训。培训科目的扩大后，机构在2002年获得了大学等级。创业型大学的变化由大学所在地杰尔市中期和长期开发概念中的而目标给与了大大的推动力，因此作为目标而确定城市从汽车业生产中心要转变为开发中心，通过其知识业的开发和多样化而协助成立有竞争力的区域化经济(Fekete、Rechnitzer，2019)。达到目标最大的方法为塞切尼·伊什特万大学的能力建设，在多个阶段后及欧盟的资金支持下完成。项目中心元素为创新能力，及研发过程有效实践所需的人士和基础设施潜力的开发，而在经济和大学分体系中确保高校的，双向知识和信息传播。

创业型大学变更后下个阶段为大学高级教育及工业合作中心（FIEK）的起步，有关的基础设施投资在2017年七月开始。项目框架内三座建筑（管理院，包裹检验实验室，

刹车片楼）的建设/基础设施的开发已经完成。在2018年完工的管理院为大学企业，及技术转让功能确保了领域，在大学内作为独立的技术中心而运转，2018年由经济开发和创新操作程序（GINOP）融资。项目基本由两大目标：1、项目最终目标为高等教育-工业合作中心（FIEK）做出一个由大学对于中小企业和符合地方经济环境的服务组合，呈现出一个包括综合的培训，及产品和组织开发服务的，可作为工业订购的产品。2、管理院（Management Campus）作为研究和培训机构 - 就大型企业和中小型企业产品开发和组织开发问题分析研究过程- 做出原创的和前沿的科学结果，之后将此结果在国际刊物上发表（Eisingerné Balassa、Rámháp，2019)。

此中心的目标就是大型企业和中小型企业组织性，及产品开发问题和烦恼的研究，以及找到科学答案。在此框架内科学分析和研就的同时可形成一个促进符合当地经济环境的，适合大学和经济企业密切长期持久合作的服务组合。管理院通过以下业务的完成而协助加强创业型大学的功能（Széchenyi István Egyetem，2018）：学生创新项目；创新服务；创新能力开发；早期阶段企业的支持（startup和衍生企业）。

创业型大学变更最后一步就是所谓的高等教育模式更新过程，其目标为建设有效的及现代化高等教育机构。其过程中模式更新机构的创立及维护权力直接从国家手中转向信托基金会。这是走向创新能力开发及商业世界开放的重要一步。从2020年八月1日起兽医大学，莫霍伊-纳吉艺术大学，米什科尔茨大学，诺伊曼·亚诺什大学，塞切尼·伊什特万大学和肖布朗大学变更为此模式运转，以及话剧和影视艺术大学作为第七所大学也加入其队伍。因此塞切尼·伊什特万大学的维护权力从国家手中由塞切尼·伊什特万大学基金会接管。"塞切尼·伊什特万大学在近几年成为本区域及匈牙利重要高等教育机构之一，具备众多的学生，基础设施投资项目，和稳定的管理。这些为继续发展确保了基础，为所有在此工作和学习者确保成为赢家的机遇"，塞切尼·伊什特万大学基金会董事会公开信中陈述到。期待此过程的结果会带来更加高效的和灵活的运转，更大的独立化，更好的开发，能够加强大学和企业合作关系，国际化及服务能力（Széchenyi István Egyetem，2020）。

# 结论，政策含义

匈牙利在完成高等教育体系的整体改革的同时还可见政策变更过程，及高校重要性的不断增长。在此一方面需要提出的是不断增长的风险资产基础的重要性，另一方面为了将地方型计划观念渗漏回原本中央化的政策中。大部分为国有的（及部分为国有的），国家及欧盟资源风险投资基金投资组合中都可以见到具备高校背景的早期阶段的

企业，以及有关的孵化器和加速器项目也绝大部分依靠着高校的资源。2021年和2027年之间预算周期有关创新目标资源利用的策略文件，智能专业化策略的准备工作，及有关区域圈内完成的探查和评估工作完成中同样由当地主要高级教育机构在《区域创新平台》框架内完成。

在区域计划中高校在新成立的经济区（超越了传统的NUTS2等级区域的计划）开发方面的开发项目计划汇编过程中也很重要。总而言之，匈牙利高等教育机构的模式变更，以及区域计划的政策及实践方面的活跃参与越来越紧密，就像埃茨科威兹和周（2018年）教授引入到科学词汇中的RIO（Regional Innovation Organizer）作用在一个创新生态系统中的重要性。创新过程中高校的参与中国式模式中呈现出良好的惯例是值得在让匈牙利考虑可适应的，无论对部分大学的管理人员，还是对匈牙利创新政策决策人员而言。这些是：

1、有关高校校办startup和衍生企业时，尤其重视符合高校长期目标的转让推出计划的策划，在此过程中主要聚焦于企业的创办人（高校）股权的保障以及长期确保稳定的运转，而且注意"耐心资本"原则。

2、地方经营者与高校创新生态体制合作的加强和加深。这不仅仅对于少数的项目，或是共同实践的建议，而是为了共同的策划及策划的实践而形成共同的策略体系。在中国高校与地方企业及各种不同的国家经营者共同设立的投资和创投基金及共同的运营尤其成功。这样的倡议不仅可以加倍国家确保的资源成本比例，而且对于一些经营者而言也是一种（资金）责任而同时加强地方合作基础的参与意愿和积极性。

中国模式中一个主要启迪就是将国际化视为主要，及支持高校和所在的地区加强其全球知名度。在这方面可见中国成功秘诀之一就是地区层和高等教育机构的合作及共同走向国际市场。匈牙利的经济区的成立（尤其是包含六个州的，拥有将近2百万人口的多瑙河以西经济区）为高校和地区经营者的合作确保了机遇和空间。这些在建立国际关系领域也值得施用，将已具备的资源和知识使用在新合作伙伴关系的建立上。

## 参考文献